\documentclass[twocolumn,preprintnumbers,amsmath,amssymb]{revtex4}
\usepackage{hyperref}
\usepackage{graphicx}% Include figure files

\begin{document}
\preprint{\vbox{\hbox{UCB-PTH-02/13}},
  \vbox{LBNL-50085}}
\title{Purely Four-dimensional Viable Anomaly Mediation}
\author{Roni Harnik}
\affiliation{Department of Physics, 
University of California, 
Berkeley, CA~~94720, USA}
\author{Hitoshi Murayama}
\author{Aaron Pierce}
\affiliation{
Department of Physics, 
University of California, 
Berkeley, CA~~94720, USA}
\affiliation{Theory Group, 
Lawrence Berkeley National Laboratory, 
Berkeley, CA~~94720, USA}
\date{\today}
\begin{abstract}
  Anomaly mediation of supersymmetry breaking solves the
  supersymmetric flavor problem thanks to its
  ultraviolet-insensitivity.  However, it suffers from two problems:
  sleptons have negative masses-squared, and there are likely bulk moduli
  that spoil the framework.  Here, we present the first fully
  ultraviolet-insensitive model of anomaly mediation with positive
  slepton masses-squared in a purely four-dimensional framework.  Our
  model is based on the additional $D$-term contributions to the sparticle 
  masses, and the conformal sequestering mechanism.
\end{abstract}
\pacs{XX.XX.xx}
\maketitle
\setcounter{footnote}{0}
\setcounter{page}{1}
\setcounter{section}{0}
\setcounter{subsection}{0}
\setcounter{subsubsection}{0}

%%%%%%%%%%%%%%%%%%%%%%%%%%%%%%%%%%%%%%%%%%%%%%%%%%%%%%%%%%%%%%%%%%%%%%%
\section{Introduction}\label{sec:intro}
The absence of superpartners with masses degenerate to those of the
Standard Model particles assures us that supersymmetry (SUSY) is
broken if it is present in nature.  However, arbitrary patterns of
supersymmetry breaking masses give flavor-changing neutral current
rates in gross violation of current experimental bounds.  This is
known as the SUSY flavor problem. 

Anomaly-mediated supersymmetry breaking \cite{RS,GLMR} represents an
attractive solution to this problem.  In anomaly mediation, the
breaking of scale invariance mediates supersymmetry breaking between
the visible and the hidden sectors. (For an expanded review of this
point, see \cite{DRED}).  As a result, the supersymmetry breaking
terms can be written entirely in terms of beta functions and anomalous
dimensions {\it at any given energy}.  This means that the SUSY 
breaking terms are independent of any flavor-dependent physics that might 
occur at high scales.  This is the anomaly-mediated solution to the SUSY 
flavor problem.  While the anomaly-mediated contribution to supersymmetry
breaking is always present, its effect can be sub-dominant.  For the
anomaly mediation to solve the SUSY flavor problem, other
contributions to supersymmetry breaking must be somehow suppressed.
In its original incarnation, this suppression was accomplished through a 
spatial separation of the visible and hidden sectors in an extra-dimension,
which was thought to have the effect of ``sequestering'' the hidden
sector \cite{RS}. 

The theory of anomaly mediation as described above suffers from two
difficulties.  First, as was shown in detail in \cite{Dine}, the
geometric separation of the visible and hidden sectors is insufficient
to sequester the hidden sector.  In fact in most cases, the exchange
of fields in the bulk supergravity multiplet is sufficient to generate
a contribution to universal scalar masses at the tree level,
completely swamping the anomaly-mediated piece.  Perhaps the simplest
counter-example to geometric sequestering appeared in the model of
radion-mediated supersymmetry breaking \cite{Chacko:2000fn}.  The
second problem facing anomaly mediation is a direct result of its ultraviolet
(UV) insensitivity. This property renders anomaly mediation an extremely
predictive framework, considered to be a desirable feature.  However,
one prediction is a negative mass squared for the sleptons, rendering
the theory phenomenologically inviable.
 
Neither of the above problems are insurmountable.  In
\cite{LutySundrum1,LutySundrum2}, it was shown that the sequestering could take
place in a completely four dimensional framework where the problem of
bulk moduli is absent, using conformal field theories.  In this case,
the sequestering remains effective as long as there are no moduli
coupled to both the hidden and the MSSM sectors.  This four
dimensional solution is reviewed in Section \ref{sec:4d}.  The problem
of tachyonic sleptons has also been overcome
\cite{RS,nondecoupling,KSS,JackJones,ViableUV}.  However, no explicit
model exists that both effectively sequesters the supersymmetry
breaking sector and provides a phenomenologically viable spectrum.  In
this paper we provide the first such model, a viable model of anomaly
mediation in four dimensions.
   
To provide the masses for the sleptons, we select the mechanism of
\cite{JackJones,ViableUV}, which gives masses through the addition of
$D$-terms.  We find this mechanism particularly attractive, as it
gives masses without sacrificing the property of UV insensitivity.  We
review this mechanism in Section \ref{sec:viable}, before moving on
and presenting our model in Section \ref{sec:model}.

\section{4D Anomaly Mediation} \label{sec:4d}
The concerns about geometric sequestering raised in \cite{Dine} can be 
addressed by utilizing the framework of Luty and Sundrum 
\cite{LutySundrum1,LutySundrum2}, in which the sequestering of anomaly mediation 
is realized in a completely four-dimensional way.  This framework is 
inspired by the AdS/CFT correspondence \cite{adscft} which conjectures 
that a 5D theory of gravity in anti-deSitter space is dual to a four 
dimensional conformal field theory. One can hope that this duality can 
be stretched further to a duality between the five dimensional brane 
world scenario and a four dimensional conformal field theory. With this 
in mind we can expect that the sequestering attempted in \cite{RS} with 
an extra dimension might be realized in a 4D model in which the hidden sector 
is conformal.  Luty and Sundrum showed this is indeed the case by looking
at 4D theory with the hidden sector being SUSY QCD with $\frac{3}{2}N_c \le
N_f \le 3N_c$ (\emph{i.e.} the theory is in the Seiberg conformal
window \cite{Seiberg}). 

The main goal of sequestering is to suppress flavor violating
operators of the form
\begin{equation}
\label{eq:contact}
\int d^4\theta \frac{c^i_j}{M_{*}^2}T^\dag_{J}T^{J}Q^\dag_{i}Q^{j}
\end{equation}
that might lead to large flavor-changing effects.  Here $T^{J}$ are
fields in the (conformal) hidden sector, $Q^{i}$ are fields in the
visible sector, and $M_*$ can be the reduced Planck scale. It is
convenient to treat such a term as a correction to the hidden sector
UV K\"ahler potential
\begin{equation}
\label{eq:UVlagranigian}
{\mathcal L}_{hid} \supset \int d^4\theta {\mathcal Z}_0 T^\dag_{J}T^{J}, \qquad
{\mathcal Z}_0=1+\frac{c^i_j}{M_{*}^2}Q^\dag_{i}Q^{j}.
\end{equation}
Below a certain energy, the gauge coupling, $g$, of the hidden sector
nears its fixed point value, $g_*$, allowing us to keep only the
leading order in $(g^2-g^2_*)$ in the renormalization group equations
for the coupling and the wavefunction renormalization.  Defining
$\gamma \equiv \partial \log{Z}/ \partial \log {\mu}$, the expansion yields:
\begin{equation}
\label{eq:RGE} 
\beta 
=\beta_{*}'(g^2-g_{*}^2), \qquad
\gamma
= \gamma_* +\gamma_*'(g^2-g_{*}^2),
\end{equation}
where $\beta_*'$ and $\gamma_*'$ are positive critical exponents,
due to the fact that they describe dynamics near an infrared fixed
point.  For simplicity we will assume Eq.~(\ref{eq:RGE}) is valid already 
at $M_*$; this assumption will not change the final result of sequestering.  
The conformal symmetry will eventually be broken by one of the fields in 
the hidden sector getting a vacuum expectation value (VEV), 
$\langle X\rangle$, away from the origin of the moduli space.

Now, we can proceed to demonstrate the sequestering.  If at the 
scale $M_{*}$, we were already at the fixed point, $g=g_*$, 
integrating Eq.~(\ref{eq:RGE}) would give 
\begin{equation}
\label{eq:fixedpt}
{\mathcal Z}(\mu)=\left( \frac{\mu}{M_*} \right)^{\gamma_*}.
\end{equation}
When considering small perturbations about the fixed point, $\Delta
g^2=g^2-g^2_*$, it is convenient to factor out the above fixed point 
running by defining
\begin{equation}
\label{eq:deltaZ}
\Delta \mbox{ln} {\mathcal Z}=\mbox{ln} {\mathcal Z}-\gamma_*\mbox{ln}\left( \frac{\mu}{M_*} \right).
\end{equation}
After integrating Eq.~(\ref{eq:RGE}) down to the scale of conformal
symmetry breaking, $\langle X\rangle$, we find that the dependence of
$\Delta \mbox{ln}{\mathcal Z}(\mu)$ on ${\mathcal Z}_0$ is highly suppressed.
This is simply a restatement of the fact that until the symmetry is
broken, the theory is nearly conformal, so that deformations
introduced at the high scale quickly become irrelevant as the theory is
driven toward its fixed point.

In particular, if we make the choice that the value of the holomorphic
gauge coupling \footnote{This is the gauge coupling whose running is
  exhausted at one loop \cite{NSVZ}, and is related to the canonical
  gauge coupling by $1/g^{2}_{can}=Re(1/g^{2}_{hol})-F \mbox{ln}{Z} /
  8 \pi^{2} - N \mbox{ln}{g_{can}^{2}}/ 8 \pi^{2}$ in the NSVZ
  scheme.} at the Planck scale is equal to its value at the fixed
point, which can be done without a loss of generality, we can write:
 \begin{equation}
\label{squester}
\Delta\ln{\mathcal Z}(\langle X \rangle )=  \left( \frac{\langle X
    \rangle}{M_*} \right)^{\beta_*'}\Delta \ln {\mathcal Z}_0.
\end{equation}
The flavor violating operators in Eq.~(\ref{eq:contact}) are therefore
power suppressed and the hidden sector is sequestered.  For details of this derivation, see References \cite{LutySundrum1,LutySundrum2}.

In \cite{LutySundrum2} a model of a conformally sequestered hidden
sector is achieved completely naturally, generating all hierarchies
dynamically. In the sections that follow, we will assume that the
hidden sector is sequestered by this mechanism, leaving anomaly
mediation as the leading contribution to visible sector soft masses.

\section{Viable Anomaly Mediation}
\label{sec:viable}
For anomaly mediation to be viable, the tachyonic sleptons must be 
eliminated. This may be accomplished by adding $D$-terms for $B-L$ and 
hypercharge.  In Reference \cite{ViableUV} the generation of a $D$-term for
$U(1)_{B-L}$ is accomplished by appealing to an extra-dimensional
framework.  We have already mentioned that the extra-dimensional
framework is problematic as a realization of anomaly mediation, so we
will have to modify this mechanism, which we do in Section
\ref{sec:model}.

The extra-dimensional set-up of \cite{ViableUV} contains three branes:
a visible sector, a sequestered sector, and a sector responsible for
the breaking of $U(1)_{B-L}$.  The $U(1)_{B-L}$ gauge field is allowed
to propagate in the bulk, but is broken at a high scale unlike in the
gaugino-mediation models \cite{gaugino}.  We now review the process by
which the $U(1)_{B-L}$ breaking sector provides a $D$-term.  The
$U(1)_{B-L}$ consists of the superpotential
\begin{equation}
\label{eqn:U1breaking}
{\mathcal W}=\lambda X (\phi \bar{\phi}- \Lambda^{2}).
\end{equation}
Here, the $\phi$ and $\bar{\phi}$ fields have $U(1)_{B-L}$ charge $+1$
and $-1$, respectively, while $X$ is $U(1)_{B-L}$ neutral.  In the 
supersymmetric limit, 
$\langle \phi \rangle=\langle \bar{\phi} \rangle=\Lambda$.  
As long as $\phi$ and $\bar{\phi}$ have different soft supersymmetry 
breaking masses, perhaps from additional Yukawa couplings not shown in
Eq.~(\ref{eqn:U1breaking}), their VEVs will shift by different amounts
after the effects of supersymmetry breaking are included \cite{HMLowEnergy}.  
The $D$-term is found to be proportional to the difference in the 
VEVs, so
\begin{equation}
D_{B-L} \sim \langle \phi \rangle^{2} -  \langle \bar{\phi}
\rangle^{2} \sim \tilde{m}^{2} - \tilde{\bar{m}}^{2}. 
\end{equation}
Generically, the $D$-term will be of the same order of the soft
masses-squared, which is the necessary condition for a viable spectrum.
The $D$-term contributes to the scalar masses-squared for all fields charged 
under the relevant symmetry, in this case, $B-L$.

The unique feature of this solution to the problem of tachyonic
sleptons is its insensitivity to physics in the ultraviolet.
More precisely, the new expressions for the soft masses:
\begin{equation}
\tilde{m^{\prime}}_{i}^{2} \equiv \tilde{m}_{i}^{2} -q_{i} D,
\end{equation}
remain on renormalization group trajectories. This can be shown by
either explicitly examining the form of the renormalization group
equations \cite{JackJones}, or by a spurion argument \cite{ViableUV}.
The spurion argument also shows clearly that soft parameters remain on
RGE trajectories even as heavy particle thresholds are crossed.

As mentioned above, $D$-terms for both $U(1)_{B-L}$ and $U(1)_{Y}$ are 
needed to generate a viable spectrum \cite{JackJones}.  However, once a 
$D$-term for $U(1)_{B-L}$ is generated, a $D$-term for $U(1)_{Y}$ is 
naturally generated by including a term for kinetic mixing in the Lagrangian,
${\mathcal L} \ni \int{d^{2} \theta \, {\mathcal W}_{B-L} {\mathcal
    W}_Y}$.  Therefore, it is sufficient to consider a mechanism to
generate $D_{B-L}$.

The ``anomaly mediation plus $D$-terms'' scenario above was originally
envisaged in an extra-dimensional framework.  If we transport this
scenario to a four-dimensional one, we require that the scale of the
$U(1)_{B-L}$ breaking be much less than the Planck scale.  Otherwise,
it is possible to generate phenomenologically dangerous contributions to 
the soft masses of the MSSM field.  To be concrete, consider the 
lepton doublet, $L$.  Operators in the K\"ahler potential of the form
\begin{equation}
\label{eq:plancksup}
\frac{(\phi L)^{\dagger} e^{qV} (\phi L)}{M_{*}^{2}}, 
\end{equation} 
will generate soft masses for the sleptons once the VEV for the
$D$-term is substituted for $V$.  A key point is that $q$ generically
will not be equal to the charge of the $L$ field.  These
contributions will take the masses off of the anomaly-mediated
trajectory, and spoil the UV insensitivity.  These operators could
arise from integrating out heavy particles charged under the
$U(1)_{B-L}$ symmetry, whose masses are at the Planck scale
\footnote{Alternatively, these could simply be additional fields
  charged under the $U(1)$ symmetry whose Yukawa couplings to $\phi$
  and $\bar{\phi}$ give those fields different soft masses.}.
Therefore, a requirement for a viable four dimensional model is that
we dynamically generate the scale of the $U(1)_{B-L}$ breaking.  We
present such a model in the next section.

\section{The Model}
\label{sec:model}
Our goal, then, is to produce a model that naturally generates a
$D$-term of the correct scale without introducing operators that spoil
the UV-insensitivity.  Looking at Sections \ref{sec:4d} and
\ref{sec:viable}, it is clear what we need to do.  Again, we will have
three sectors: a visible sector containing the MSSM, a nearly
conformal hidden sector, and a $U(1)_{B-L}$ breaking sector.  The
nearly conformal sector can be neatly lifted from Reference
\cite{LutySundrum2}.  Our task is to generate a superpotential similar
to that in Eq.~(\ref{eqn:U1breaking}) dynamically.  This will generate
the $D$-terms, allowing for a phenomenologically viable spectrum.

The sector that we use to break the $U(1)$ gauge symmetry and produce
the $D$-term is given in Table~\ref{tab:content}.  It is a theory with
a ``strong'' $Sp(2)$ gauge group and 3 flavors, corresponding to 6
fundamentals, which we denote by $Q$.  The theory exhibits an $SU(6)$
flavor symmetry.  We gauge the $SU(4) \times SU(2) \times U(1)$
subgroup of the flavor symmetry, and this $U(1)$ is the $U(1)_{B-L}$
that develops a $D$-term \footnote{A similar attempt using $Sp(1)$
  with 2 flavors runs into the difficulty that the mesons have
  identical charges under the $SU(3) \times U(1) \subset SU(4)$ flavor
  symmetry, so it is not entirely clear that a $D$-term is generated.}.
To cancel the anomalies for the $SU(4) \times SU(2) \times U(1)$
symmetry, we introduce four copies of anti-flavors $\bar{Q}$
transforming as a $\bar{4}$ under a flavor $SU(4)_{g}$ global symmetry
but not charged under the strong $Sp(2)$ gauge group.

\begin{table}[htbp]
  \begin{center}
    \begin{tabular}{|c|c|c|c|}
      \hline
      & $Sp(2)$ & $SU(6)\supset SU(4)\times SU(2) \times U(1)$
      & $SU(4)_{g}$\\ \hline
      $Q$ & 4 & $6=(4,1)_{1} + (1,2)_{-2}$ & 1\\
      $\bar{Q}$ & 1 & $\bar{6}=(\bar{4},1)_{-1} + (1,2)_{2}$ & 4\\
      \hline
      $M$ & 1 & $15=(6,1)_{2} + (4,2)_{-1} + (1,1)_{-4}$ & 1 \\ \hline
    \end{tabular}
    \caption{Particle content of the U(1) breaking sector and 
        associated mesons.}
    \label{tab:content}
  \end{center}
\end{table}

The superpotential of the model is
\begin{equation}
  \label{eq:W}
  W = \frac{1}{M_*} (Q^i Q^j) \bar{Q}_i \bar{Q}_j + \frac{1}{M_*^3} Q^6 + 
	\frac{1}{M_{*}^{3}} \bar{Q}^{6},
\end{equation}
which is the most general superpotential up to order $1/M_{*}^{3}$ consistent with the symmetries described above.  The
indices run from 1 to 6 over all three flavors.  However, the terms
do not have to be fully $SU(6)$ symmetric, but symmetric only under
$SU(4)\times SU(2) \times U(1)$, {\it i.e.}\, different coefficients
for independent invariants are allowed in the last two terms.

The dynamics of the $Sp(2)$ gauge group leads to the quantum modified
constraint \cite{IP}
\begin{equation}
  {\rm Pf}\ M = \Lambda^6,
\end{equation}
where the meson is made of quarks $M^{ij} = Q^i Q^j$.  Because the
quarks are in $(4,1)_{1} + (1,2)_{-2}$ representations, the mesons
transform as $(6,1)_{2} + (4,2)_{-1} + (1,1)_{-4}$.  The quantum
modified constraint can be satisfied consistently with vanishing $SU(4)
\times SU(2) \times U(1)$ $D$-terms with
\begin{equation}
  M_{(6,1)_2} = (\sqrt{2}\ V^2, 0, 0, 0, 0, 0), \qquad
  M_{(1,1)_{-4}} = V^2,
  \label{eq:mesonVEVs}
\end{equation}
where $V^6 = \Lambda^6$.  The first meson VEV breaks $SU(4)$ to
$Sp(2)$ (or equivalently $SO(6)$ to $SO(5)$), and both of them break
$U(1)$.  Most of the mesons in the $(6,1)_{2} + (1,1)_{-4}$
representations are eaten by the broken gauge multiplets, and one of
them is eliminated due to the quantum modified constraint.  The
$(4,2)_{-1}$ mesons do not acquire mass because of the ``accidental''
$SU(6)$ symmetry in $Sp(2)$ dynamics and are pseudo-Nambu--Goldstone
fields.  The second term in Eq.~(\ref{eq:W}) then breaks the
``accidental'' $SU(6)$ and gives the $(4,2)_{-1}$ mesons masses of
order $\Lambda^4/M_*^3$.  On the other hand, the first term in Eq.~(\ref{eq:W})
gives mass of order $\Lambda^2/M_*$ to all of the anti-flavors.  As a
result, all fields that we have introduced acquire supersymmetric
masses.  There are no new light fields in our model.

We claim that, in the presence of soft supersymmetry breaking induced
by anomaly mediation, this model develops a $D$-term for the $U(1)$.  

Unfortunately, this model is incalculable because the mesons become
composite and condense at the same scale, and we do not know if the
description in terms of meson degrees of freedom is appropriate to
work out the soft supersymmetry breaking effects.  Of course, once
the dynamical degrees of freedom are identified, anomaly mediation gives
a unique prediction for their soft parameters.  However, we do not know
if mesons can be regarded as dynamical degrees of freedom or composite
order parameters at this stage.

We make this model calculable, in the same spirit as in \cite{HMSO10},
by deforming the theory while remaining in the same universality
class.  The model becomes calculable if two energy scales are
separated: the scale where mesons become composite $\Lambda$, and the
scale where $U(1)$ is broken $V$.  We identify two deformations which
achieve $\Lambda \gg V$ or $\Lambda \ll V$.  We will find that a
$D$-term of the correct order of magnitude is successfully generated
in either case.  We view this as compelling evidence for the
generation of a $D$-term when $\Lambda \sim V$, which is the actual
situation in the model we have presented.

\subsection{$\Lambda \gg V$}

The first limit, $\Lambda \gg V$, is achieved by introducing an
additional flavor, $Q_{7}, Q_{8}$.  No corresponding anti-flavor is necessary 
as the additional flavor is not charged under any gauge group except the
strong $Sp(2)$.  The extra flavor is massive.  When its mass $m \gg
\Lambda$, we can integrate out the extra flavor first, and the theory
goes back to the original model.  On the other hand, by taking $m \ll
\Lambda$, the $Sp(2)$ dynamics changes at the dynamical scale.  The
theory has 4 flavors and hence confines with a dynamical
superpotential \cite{IP}.  Together with the mass term,
\begin{equation}
  \Delta W = - m {\mathcal M}^{78} + \frac{\mbox{Pf}\ {\mathcal M}}{\Lambda^{5}}.
  \label{eq:W2}
\end{equation}
The meson ${\mathcal M}$ includes the 4th flavor, while we keep the
notation $M$ for the first three flavors.  Below the scale $\Lambda$,
the confined mesons are dynamical degrees of freedom, as suggested by
the non-trivial anomaly matching conditions.

At the scale $V = (m\Lambda^5)^{1/6}$, we solve for ${\mathcal M}^{78}$
and find
\begin{equation}
  \mbox{Pf}\ M = m \Lambda^5 = V^6.
\end{equation}
which sets $\Lambda \gg V$ because $m \ll \Lambda$.
The mesons acquire expectation values in the same fashion as in
Eq.~(\ref{eq:mesonVEVs}), breaking $U(1)$ and $SU(4) \rightarrow
Sp(2)$. 

The superpotential of Eq.~(\ref{eq:W}) becomes Yukawa
couplings among mesons and anti-flavor quarks
\begin{equation}
  W = \frac{1}{M_*} M^{ij} \bar{Q}_i \bar{Q}_j + \frac{1}{M_*^3} M^3.
\end{equation}
Once mesons acquire VEVs, the first term gives the anti-flavors masses, and 
the second term gives masses to the $(4,2)_{-1}$ mesons.  The size of Yukawa
couplings can be seen by scaling the meson fields to canonical normalization 
$M \approx \Lambda \hat{M}$ up to unknown $O(1)$ constants.

The main point of this deformation is that physics between the
compositeness scale, $\Lambda$, and the $U(1)$-breaking scale, $V$, is
given in terms of composite mesons and their soft supersymmetry
breaking can be obtained by the standard formulae of anomaly
mediation.  Yukawa couplings from the non-renormalizable
superpotential are suppressed by $\Lambda/M_*$ and we ignore them.
Then the only contribution to their soft masses come from $SU(4)
\times SU(2) \times U(1)$ gauge interactions.  The relevant soft
masses for the mesons are determined by the expression
$\tilde{m}^{2} =(-\dot{\gamma}/4) |m_{3/2}|^{2}$, where $\gamma \equiv
(\partial \log{Z} / \partial \log \mu )$.  Calculating in the gauge
theory, we find:
\begin{eqnarray}
\label{eqn:mesonmasses}
  \tilde{m}^2_{(6,1)_{2}} &=& \frac{|m_{3/2}|^{2}}{(16 \pi^{2})^{2}}
  (20 g_{4}^{4}-384 g_{1}^{4}) ,\\
\label{eqn:mesonmasses2} 
  \tilde{m}^2_{(1,1)_{-4}} &=& \frac{|m_{3/2}|^{2}}{(16 \pi^{2})^{2}}
  (-1536 g_{1}^{4}). 
\end{eqnarray}
In the presence of soft terms, we have to re-minimize the potential
obtained from the superpotential Eq.~(\ref{eq:W2}). The vacuum
configuration for the new potential shifts slightly from that in
Eq.~(\ref{eq:mesonVEVs}).  And the $D$-term, which is given in terms
of the shifts in the meson VEVs, can be given by:
\begin{equation}
D=8g_{1} (\delta  M_{(6,1)_2} - \delta M_{(1,1)_{-4}})=
\frac{\tilde{m}^2_{(1,1)_{-4}}-\tilde{m}^2_{(6,1)_{2}}}{6 g_{1}}. 
\end{equation}
As an aside, we note that the potentially large $A$-terms that result
from inserting powers of the chiral compensator in Eq.~(\ref{eq:W2}),
do not contribute to the $D$-term.  We have checked that this is the case 
using the methods of \cite{HMLowEnergy}.  Now, substituting in the expressions
of Eqs.~(\ref{eqn:mesonmasses}) and (\ref{eqn:mesonmasses2}), the final value 
for the $D$-term in this case are given by:
\begin{equation}
D=-\frac{|m_{3/2}|^{2}}{(16 \pi^{2})^{2}}\frac{192 g_{1}^{4} +
  \frac{10}{3} g_{4}^{4}}{ g_{1}}, 
\end{equation}
so a $D$-term of the correct order of magnitude is successfully
generated in this limit.

\subsection{$\Lambda \ll V$}

The second limit $\Lambda \ll V$ is achieved by introducing an
additional flavor, $Q_{7}, Q_{8}$, and extending the strong gauge
group to $Sp(3)$.  Because we are extending the gauge group to
$Sp(3)$, now six copies of the anti-flavors are necessary.  We assign 
the extra fourth flavor the superpotential:
\begin{eqnarray}
  \Delta W = \lambda Y (Q^7 Q^8 - v^2)
  + \lambda' Y_{i\alpha} Q^i Q^\alpha.
  \label{eq:W3}
\end{eqnarray}
Here, $i=1, \cdots, 6$ and $\alpha=7,8$. 
We also extend the original superpotential by letting the indices in Eq. (\ref{eq:W}) run over $1,\ldots,8$.
When $v \gg \Lambda$, the
extra flavor condenses and breaks the gauge group $Sp(3) \rightarrow
Sp(2)$, while it is eaten by the gauge multiplet.  The extra color
degrees of freedom of the remaining 3 flavors are made massive by the
second term in Eq.~(\ref{eq:W3}).  The first term in Eq.~(\ref{eq:W})
makes the extra anti-flavors massive.  Therefore the theory reduces to
the original model.

When $v \ll \Lambda$, on the other hand, it yields a calculable model
of $D$-term generation with the scale of $U(1)$ breaking $V \gg
\Lambda$.  The soft masses can be calculated from anomaly mediation,
the corresponding shift in the VEVs can be computed, and the $D$-term
can be identified.  The quick way to see that the model is still in
the same universality class is by using the dynamical superpotential.
Together with the quantum modified constraint,
\begin{equation}
  \Delta W = \lambda Y ({\mathcal M}^{78} - v^2) 
  + \lambda' Y_{i\alpha} {\mathcal M}^{i\alpha}
  + X ({\rm Pf}\ {\mathcal M} - \Lambda^8).
  \label{eq:W4}
\end{equation}
The meson ${\mathcal M}$ includes the 4th flavor, while we keep the
notation $M$ for the first three flavors.  We can immediately conclude
that ${\mathcal M}^{78} = v^2$, ${\mathcal M}^{i\alpha} = 0$.  Therefore,
${\rm Pf}\ {\mathcal M} = ({\rm Pf}\ M) {\mathcal M}^{78} = ({\rm Pf}\ M)
v^2$.  The last term in the superpotential of Eq.~(\ref{eq:W4}) then 
determines $({\rm Pf}\ M) = \Lambda^8/v^2$.  Therefore the meson VEV 
breaks $SU(4) \times SU(2) \times U(1) \rightarrow Sp(2) \times SU(2)$ 
just as before.  What is important is the scale of the VEVs: when 
$v \ll \Lambda$, ${\rm Pf}M \gg \Lambda^6$, and hence the meson VEVs 
corresponding to the $U(1)$ breaking scale, $V$, 
can be regarded as classical expectation values of quark fields.  Therefore 
we can calculate the effects of soft parameters using the anomaly mediation
formula for quark fields instead of composite fields.

At the scale $(\Lambda^4/v)^{1/3}$, we look at classical flat
direction 
\begin{equation}
  Q = \left( \begin{array}{cccccc|cc}
      V&0&0&0&0&0&0&0\\
      0&V&0&0&0&0&0&0\\
      0&0&V&0&0&0&0&0\\
      0&0&0&V&0&0&0&0\\
      0&0&0&0&V'&0&0&0\\
      0&0&0&0&0&V'&0&0
    \end{array} \right).
\end{equation}
In this notation, the $Sp(3)$ gauge group acts from the left, while
the $SU(4) \times SU(2) \times U(1)$ gauge groups act from the right
on the quark multiplets.  Because $V,V' \sim (\Lambda^4/v)^{1/3} \gg
\Lambda$, it is a conventional Higgs mechanism that breaks $Sp(3)
\times SU(4) \times SU(2) \times U(1) \rightarrow Sp(2) \times SU(2)$.
Out of $6\times 8=48$ components of the quark multiplets,
$(21+15+3+1)-(10+3)=27$ components are eaten.  The 12 components in
the fourth flavor become massive together with $Y_{i\alpha}$ due to
the second term in Eq.~(\ref{eq:W3}).  The $(4,2)$ components acquire mass
from the second term in Eq.~(\ref{eq:W}).  This leaves only one light
degree of freedom, namely the direction of $V$ itself.  We, however,
leave the possibility that $V \neq V'$ in the above expression.  This
is justified when the $U(1)$ coupling is small.  The $D$-flatness is
imposed because the $D$-non-flat direction acquires a mass of order
$g_{1}V$.  When $g_{1} \ll 1$ we can consider a low-energy effective theory
where all particles of mass $\sim V$ are integrated out but those of
mass $g_{1} V$ are not.  We can minimize the potential with respect to $V$
and $V'$ later.

Along this flat direction, we need to consider the instanton effect of
partially broken gauge groups \cite{CM}.  Because $\pi_3 ((Sp(3)\times
SU(4) \times SU(2) \times U(1))/(Sp(2) \times SU(2))) = {\mathbb Z}$,
there are instanton effects that are present in the $Sp(3)\times SU(4)
\times SU(2) \times U(1)$ which are not part of the low-energy $Sp(2)
\times SU(2)$ theory.  Such effects have to be included in writing
down the low-energy theory.  In our case, it is
\begin{equation}
  {\mathcal M}^{78} = \frac{\Lambda^8}{V^4 V^{\prime 2}}.
\end{equation}
This makes the first term in the superpotential Eq.~(\ref{eq:W3})
\begin{eqnarray}
  W = \lambda Y \left(\frac{\Lambda^8}{V^4 V^{\prime 2}} - v^2\right)
  \label{eq:W5}
\end{eqnarray}
while the second term had already been used to integrate out the
fourth flavor.

Therefore, the low-energy effective theory we need to solve is given by
the directions $V$ and $V'$ of the elementary quarks together with the
superpotential Eq.~(\ref{eq:W5}) and soft supersymmetry breaking
effects calculated in the quark language.

Calculating in this framework, we find that the $D$-term can be given
in terms of the quark soft masses as
\begin{equation}
D=\frac{\tilde{m}^{2}_{(1,2,-2)}- \tilde{m}^{2}_{(4,1,1)}}{3 g_{1}}
\end{equation}
The quark soft masses themselves arise from anomaly mediation and are given by:
\begin{eqnarray}
 \tilde{m}^2_{(4,1,1)} &=& \frac{|m_{3/2}|^{2}}{(16
   \pi^{2})^{2}}(\frac{45}{4}g_4^{2} -144 g_1^{4})+m_{Sp(3)}^{2},\\ 
 \tilde{m}^2_{(1,2,-2)} &=& \frac{|m_{3/2}|^{2}}{(16
   \pi^{2})^{2}}(-576 g_{1}^{4}) +m_{Sp(3)}^{2}. 
\end{eqnarray}
Where we have denoted the universal contribution coming from the
$Sp(3)$ gauge interactions as $m_{Sp(3)}^{2}$.  Note that the $SU(2)$
theory is conformal at one-loop, so its coupling does not contribute
to $\tilde{m}^2_{(1,2,-2)}$.  The final expression for the $D$-term in
this case is then
\begin{equation}
D=-\frac{|m_{3/2}|^{2}}{g_{1} (16 \pi^{2})^{2}} (\frac{15}{4}
g_{4}^{4} + 144 g_{1}^{4}). 
\end{equation}
We view the fact that a $D$-term is generated in both deformations
(with the same sign, no less) as compelling evidence for existence of
a $D$-term in the case $\Lambda \sim V$, which is the situation in our
model for $D$-term generation.

\subsection{UV Insensitivity}
Let us return to the questions of UV-insensitivity.  There might be a
concern that additional contributions to the soft masses of the MSSM
fields might arise from integrating out other heavy fields charged
under the $U(1)$.  However, because of the charges of the $Q$'s under
$Sp(2)$ and $SU(4) \times SU(2) \times U(1) \subset SU(6)$, it is
impossible to construct a non Planck-suppressed invariant with these
fields that could couple to the MSSM.  As long as the dynamically
generated scale is much less than the Planck scale, these
contributions are small.  One might imagine generating
non-Planck-suppressed operators if there were further fields charged
under the $Sp(2)$ gauge symmetry.  However, this would upset the
$Sp(2)$ dynamics that breaks $U(1)$.  So, once we have specified our
mechanism for the breaking of $U(1)$, there is no room for additional
contributions to the soft-masses that will displace them from their
renormalization group trajectories.

\section{Phenomenology}
The spectrum of the superpartners should be nearly identical to that
presented in references \cite{JackJones,ViableUV}, as we have merely
presented a field theoretic realization of that scenario.  Electroweak
symmetry breaking has not been fully investigated in this scenario,
and it would be interesting to explore the possibility of
incorporating the NMSSM into this framework.  We note that the
sequestering mechanism of \cite{LutySundrum2} is not perfect, and it is
possible that interesting flavor changing effects could be observed
soon in the context of this framework.

There are no new light fields in our model.  All fields receive
supersymmetric masses.  That said, we prefer that dynamically
generated scale not be too low.  The mesons transforming as
$(4,2)_{-1}$ only get a mass from the last term of Eq.~(\ref{eq:W}),
so their masses are expected to be $m_{(4,2)_{-1}} \sim \Lambda
(\Lambda/M_{*})^{3}$.  In the model as written, these fields are
absolutely stable, so there could be a concern that they might
over-close the universe.  However, as long as $\Lambda > 10^{16}$ GeV,
they place no constraint on the model above and beyond the usual
constraint coming from not producing an over-abundance of gravitinos 
\cite{Kawasaki:1994af}.  This value for $\Lambda$ is sufficiently small 
that the Planck-suppressed contributions to the K\"ahler potential of the 
form of Eq.~(\ref{eq:plancksup}) are negligible.  Of course, lower values
for $\Lambda$ would be possible for smaller values of the re-heating
temperature.  If rather than being absolutely stable, these particles
decayed through sufficiently Planck suppressed operators, it is
possible that their lifetime could be roughly the age of the universe.
In this case, they could provide a mechanism to explain the existence
of ultra-high energy cosmic rays.
 
It is interesting to note that this mechanism accommodates the
see-saw mechanism \cite{seesaw} for neutrino masses.  We can add the 
couplings $W \supset (N^{2} M_{(1,1)_{-4}})/M_{*} + h L N H_{u}$  to the 
superpotential (where we normalize our $U(1)$ so that $N$ has charge 2).  
Then, as the mesons acquire VEVs, the first term provides a mass of the 
right scale for atmospheric neutrino masses, assuming the Dirac masses 
are roughly the top mass, as suggested by the $SO(10)$ mass relation.  
Note that there are corrections to the RGE trajectories at higher orders 
in this case, but the leading order result remains unaffected \cite{ViableUV}.
  In particular, one expects that the operator 
\begin{equation}
  {\mathcal K} \ni \frac{|h|^{2}}{16 \pi^{2}} 
  \frac{(L^{\dagger} e^{q_{L} V} L) (M^{\dagger} e^{q_{M} V} M)}
  {M_{*}^{2} M_{R}^2},
\end{equation}
will be generated in the K\"ahler potential by integrating out the $N$
field.  Here, $M_{R}\sim M/M_*=\Lambda^2/M_*$ is the right handed
neutrino mass.  Inserting the vacuum expectation values for the mesons
and the $D$-term, this represents a perturbation away from the anomaly
mediated trajectory at the percent level, and it is not entirely
automatic that lepton flavor-changing neutral current effects will be
safe.  On the other hand, this might lead to interesting signals, for
example, in processes such as $\mu \rightarrow e \gamma$.

Another option to get neutrino masses is to forbid the usual $L H_{u} N$ 
term in the superpotential, but include the non-renormalizable operator 
$L H_{u} N / M_{*}$ in the K\"ahler potential.  After inserting appropriate 
powers of the chiral compensator, and expanding 
$\langle \phi^{\dagger} \rangle = 1+ \bar{m}_{3/2} \bar{\theta}^{2}$, a 
term $L H_{u} N$ is generated in the superpotential, with Yukawa coupling 
of order $m_{3/2}/M_{*}$, allowing for Dirac neutrinos of the correct 
scale \cite{AdditionalNu}. 

\section{Conclusion}
\label{sec:conclusion}
We have presented the first fully viable model of anomaly mediation.
Using the sequestering mechanism of Luty and Sundrum, we are assured
that anomaly-mediated supersymmetry breaking dominates, thereby
solving the flavor problem.  Supplementing their sequestering
mechanism with a mechanism for generating $D$-terms, we have given an
example of a UV-insensitive four dimensional supersymmetry breaking.

\acknowledgments{ This work was supported in part by the DOE Contract
  DE-AC03-76SF00098 and in part by the NSF grant PHY-0098840}

\end{document}